\documentclass[lettersize,journal]{IEEEtran}
\usepackage{amsmath,amsfonts}
\usepackage{algorithmic}
\usepackage{algorithm}
\usepackage{array}
\usepackage[nolist]{acronym}
\usepackage[caption=false,font=normalsize,labelfont=sf,textfont=sf]{subfig}
\usepackage{textcomp}
\usepackage{stfloats}
\usepackage{url}
\usepackage{verbatim}
\usepackage{graphicx}
\usepackage{tikz}
\usetikzlibrary{positioning}
\usepackage{cite}
\usepackage{siunitx}
\usepackage{lipsum}
\usepackage{hyperref}
\usepackage{tablefootnote}
\begin{document}

\begin{acronym}
\acro{hbm}[HBM]{high-bandwidth memory}
\acro{sram}[SRAM]{static random-access memory}
\acro{ue}[UE]{user equipment}
\acro{gnb}[gNB]{next-generation Node~B}
\acro{bleu}[BLEU]{bilingual evaluation understudy}
\acro{ai}[AI]{artificial intelligence}
\acro{rrc}[RRC]{radio resource control}
\acro{rse}[RSE]{\acs{rrc} sequence engine}
\acro{rca}[RCA]{root-cause analysis}
\acro{rtt}[RTT]{round-trip time}
\acro{ie}[IE]{information element}
\acro{llm}[LLM]{large language model}
\acro{sft}[SFT]{supervised fine-tuning}
\acro{asn1}[ASN.1]{Abstract Syntax Notation One}
\acro{f1ap}[F1AP]{F1 application protocol}
\acro{xnap}[XnAP]{Xn application protocol}
\acro{ngap}[NGAP]{NG application protocol}
\acro{nas}[NAS]{non-access stratum}
\acro{e1ap}[E1AP]{E1 application protocol}
\acro{gnbcucp}[gNB-CU-CP]{gNB central unit-control plane}
\acro{gnbcuup}[gNB-CU-UP]{gNB central unit-user plane}
\acro{gnbdu}[gNB-DU]{gNB distributed unit}
\acro{amf}[AMF]{access and mobility management function}
\acro{ttlt}[TTLT]{time to last token}
\acro{em}[EM]{exact match}
\acro{kpi}[KPI]{key performance indicator}
\acro{3gpp}[3GPP]{3rd Generation Partnership Project}
\end{acronym}

\title{Sequence Models for Layer-3 Protocol Emulation}

\author{Alix Jeannerot, Petko Petkov, Alvaro Valcarce Rial 
\thanks{Alix Jeannerot and Alvaro Valcarce Rial are with Nokia Bell Labs, Massy, France; Petko Petkov is with INSAIT, Sofia University “St. Kliment Ohridski”, Bulgaria; parts of this work were carried out while Petko Petkov was at Nokia Bell Labs, Massy, France. The authors would like to thank Pavan Koteshwar Srinath and Istvan Kovacs for their comments. Generative-AI tools have been used to improve grammar and wording of the paper.}}

\maketitle
\begin{abstract}
This article investigates whether Layer-3 radio-protocol behavior can be represented by compact sequence models suitable for deployment inside the RAN.
We introduce the \ac{rse}, a hybrid architecture in which a sequence model predicts protocol-dependent message structure while deterministic components retain control over security-sensitive or configured fields and over transport containers.
Using NR protocol traces, we show that protocol-aware tokenization, cross-stack context, and explicit placeholders matter more than general-sequence model scale.
A purpose-built 11M-parameter Mamba engine achieves 0.90 exact match on the evaluated \ac{gnb}-side test set with a median generation latency of 115 ms, sufficient for some timers of Layer-3, outperforming a fine-tuned 0.6B-parameter model.
We discuss applications in testing, simulation, deployment specialization, and future trainable 6G control planes, as well as the validation, latency, robustness, and security challenges that remain before operational use.
\end{abstract}



\section{Introduction}
The \ac{rrc} protocol~\cite{etsiTS1383312025} is a Layer-3 control-plane protocol between the \ac{ue} and the \ac{gnb}. Among others, it governs the connection and release of \acp{ue} to \acp{gnb}, the broadcast of system information, the radio-bearer management, and the mobility procedures. A correct and efficient \ac{rrc} implementation is therefore essential to network stability and to the user experience.

Like other cellular protocols, \ac{rrc} is specified by the \ac{3gpp}, where working groups collect contributions from vendors, operators, and research organizations, debate them in successive meetings, and freeze the agreed procedures into a release of the specification. The resulting document contains procedure descriptions and message definitions that each vendor then implements. This process has well-known drawbacks. Because every change must accommodate a global set of stakeholders and use cases, the protocol grows by accumulation, the specification spanning thousands of pages with mandatory procedures interleaved with optional ones. As a result, only a fraction of the standardized procedures are effectively implemented and productized. Furthermore, customization of the protocol to optimize a specific deployment is impossible unless it was anticipated in the specification. The cadence of meetings and the subsequent implementation cycle also mean that the control plane evolves slowly: a feature agreed in one release typically takes years to appear in commercial networks. Lastly differences in interpretation of the same text further yields vendor-specific stacks that are difficult to compare, test, or replace, making interoperability certification costly. 

These limitations have motivated alternative ways of realizing and implementing standardized protocols. Recent work~\cite{liuLLMBasedEmulationRadio2026} showed that a fine-tuned \ac{llm} can partially emulate \ac{rrc}: real and simulated network traces are presented to a Llama model, which produces the corresponding response. The key insight is to treat \acs{asn1} \ac{rrc} messages as text, and therefore as sequences of predictable tokens. The results are encouraging, but not yet practical. Inference on an 1-billion parameter transformer takes a few seconds, far beyond the latency expected for Layer~3, and the generated message is not guaranteed to be the one the \ac{ue} expects. 

This article treats \ac{rrc} as a domain-specific language on which the techniques developed for language models can be applied, and calls the sequence model that results from this treatment an \ac{rse}. We ask how far an \ac{rse} can be pushed toward the latency and reliability of rule-based \ac{rrc} stacks. We follow two routes: fine-tuning a compact Qwen \cite{yangQwen3TechnicalReport2025} model, and training a custom Mamba \cite{guMambaLinearTimeSequence2024a} architecture from scratch. Both rely on a protocol-aware tokenizer and on curated inputs and outputs. An important takeaway is that, for an \acs{ai}-based protocol, message \emph{content} and protocol \emph{logic} should be decoupled to obtain a high accuracy. By doing so, the \ac{rse} is not asked to infer operator-provisioned or security-sensitive fields and focuses on creating the correct message with appropriate \acp{ie}. Although we focus on \ac{rrc}, the same methods apply to other Layer-3 protocols in the 5G stack or in other radio technologies.

The remainder of this article is organized as follows. Section~\ref{sec:motivations} motivates protocol emulation and reviews related work. Section~\ref{sec:system} describes where the \ac{rse} sits in a disaggregated \ac{gnb} and how traces are turned into training conversations. Section~\ref{sec:models} presents the two modeling routes. Experimental results follow in Section~\ref{sec:results}.

\section{RRC Sequence Engines: Motivation and Existing Work}
\label{sec:motivations}
\subsection{Motivation}
An \ac{rrc} layer based on an \ac{rse} opens applications that are difficult to address with conventional, rule-based stacks. By emulating the control plane, it approximates observed behavior from traces, and can be deployed where a full stack is unavailable, impractical, or undesirable.

A first class of applications concerns \textbf{operational troubleshooting}. By replaying captured traces through an \ac{rse}, engineers can isolate whether an observed anomaly originates from \ac{rrc} logic itself or from interactions with neighboring protocols. Because the model is trained on network traces, it can also serve as a behavioral reference against which another implementation can be compared, supporting \ac{rca} without access to the vendor's source code. Closely related is what-if analysis: once a faithful \ac{rse} is available, protocol parameters or message sequences can be perturbed in a controlled setting, so that counterfactual scenarios can be studied without a full network simulation or a change to live equipment.

\Acp{rse} are also valuable for \textbf{interoperability testing}. An engine trained on a specific vendor's traces can be used as a black-box peer, so that devices are exercised against realistic signaling without requiring a matching commercial stack. 

Another interesting areas of application are \textbf{simulation and digital twins}. Network simulators, like ns-3 \cite{patricielloE2ESimulator5G2019}, which often omit or oversimplify Layer~3, could instead use an \ac{rse} trained on real traces and gain protocol fidelity without the engineering cost of implementing the standard or needing access to operator-owned protocol traces. That trade-off is particularly attractive for research platforms and early integration testing, where realism of the exchanged messages matters more than completeness of every procedure.

\textbf{Site-specific customization} is another compelling use case. Fine-tuning or distilling an \ac{rse} on traces from a particular deployment yields a lean protocol implementation tailored to local conditions, again without access to the vendor's codebase. The resulting \ac{rse} need only cover the procedures actually observed in the field. In a fixed wireless access deployment, for example, handover support may be entirely absent from operational traces and can therefore be omitted, reducing complexity and resource consumption while preserving the procedures that matter. Another example is non-terrestrial communications where content of message could be revisited to minimize the number of message exchange (as the \ac{rtt} is usually high).

Looking further ahead, protocol emulation is a stepping stone towards \textbf{protocol emergence} \cite{motaEmergenceWirelessMAC2021a}. If \ac{ue}-side and \ac{gnb}-side \acp{rse} interact through realistic message exchanges, they form a closed loop in which control-plane behavior can be shaped under deployment constraints, rather than solely by a fixed specification. \Acp{rse} provide a parameterized, trainable policy that can be optimized through supervised or reinforcement learning or jointly with AI models in other layers. With appropriate reward of objective, this could lead to the discovery of more efficient signaling for the environment in which they operate. That longer-term perspective aligns with the AI-RAN vision: intelligence is embedded throughout the stack, and cross-layer and multi-agent optimization produces networks that adapt to their surroundings.

\subsection{Related Work}
Two main lines of research currently explore the usage of \acp{llm} for cellular networks. The first is analytical: in~\cite{sanaReasoningLanguageModels2025a}, reasoning models produce structured multi-step explanations of network traces. A curated troubleshooting corpus and a mix of supervised fine-tuning and reinforcement learning are used to produce domain-adapted models that outperform general-purpose reasoning methods on diagnostic accuracy. Similar applications revolve around specification comprehension, question answering over \ac{3gpp} documents, and operational troubleshooting~\cite{maatoukLargeLanguageModels2025b}. In all of these cases the model reads traces or standards and produces explanations or error reports but does not emit the protocol messages that the network would actually exchange.

The second line of research seeks to apply pre-trained \acp{llm} directly or indirectly within the network. A recent example is \textit{The AI Telco Engineer} \cite{aoudiaAutonomousDiscoveryWireless2026}, which uses language models to discover and implement wireless algorithms. In contrast, in~\cite{liuLLMBasedEmulationRadio2026}, it is proposed to directly integrate a language model in the network stack by replacing the whole \ac{rrc} protocol implementation. By considering \ac{rrc} messages as a domain-specific language, a Llama decoder was fine-tuned with low-rank adaptation on a multi-vendor corpus of real 4G and 5G traces, recast as uplink-downlink question-answer pairs. The resulting \ac{rse} attains high syntactic conformance and a high cosine similarity to ground-truth messages, showing that a language model can reproduce control-plane procedures from traces. Nonetheless, several obstacles remain. Inference is far slower than Layer-3 requirements (median generation times of several seconds). Moreover, a high cosine similarity does not guarantee an accurate message, as two different message types that share vocabulary can appear close in embedding space yet trigger different actions at the receiver. Finally, the \ac{rse} sees only \ac{rrc} history, although neighboring protocols often trigger a decision. The remainder of this article addresses these gaps with compact \acp{rse}, a domain-adapted tokenizer, placeholders that isolate unpredictable fields, and cross-protocol context, with the aim of bringing \acp{rse} as close as possible to real-time, rule-based stacks.

\section{System Model and Data Representation}
\label{sec:system}
\subsection{Where the Sequence Engine Sits}
\begin{figure}
  \centering
  \includegraphics[width=\linewidth]{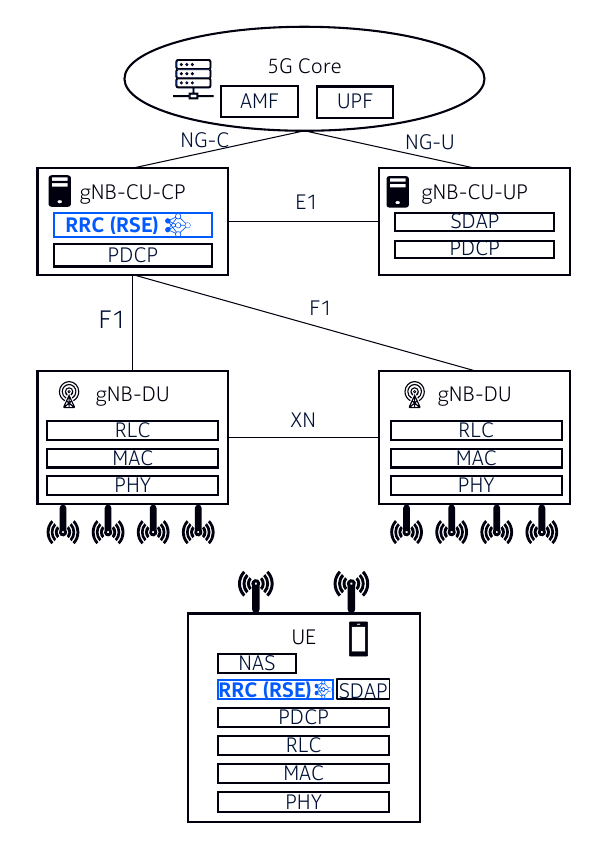}
  \caption{Depiction of the different components and layers of a cellular network and how they interact with the \ac{gnbcucp}. The uplink and downlink \ac{rrc} messages transit through the \ac{gnbdu} (without any modification) via the F1 interface before reaching the \ac{gnbcucp} or \ac{ue}.}
  \label{fig:cu_and_friends}
\end{figure}

The \acp{rse} aim to emulate the \ac{rrc} in the \ac{gnbcucp} and in the \ac{ue}. As shown in Figure~\ref{fig:cu_and_friends}, it does not replace the physical layer, the medium-access layer, or the transport of messages. At the \ac{gnb}, uplink \ac{rrc} messages arrive at a \ac{gnbdu} and are forwarded transparently to the \ac{gnbcucp} via F1 \texttt{UlRrcMessageTransfer} containers; downlink messages travel in the opposite direction via \texttt{DlRrcMessageTransfer} containers.

To capture all the context required for the \ac{gnb}-side \ac{rse}, it is necessary to feed it with messages coming from other protocols in addition to the \ac{rrc} messages coming from the \ac{ue}. In particular, \ac{nas}, \ac{f1ap}, \ac{xnap}, \ac{ngap}, and \ac{e1ap} messages should be included. These other messages, referred to below as \textit{external} protocol messages (as they are external to \ac{rrc}), carry the signaling between the \ac{gnbcucp} and the rest of the radio access network or the core. Because these protocols are confined to the network side, the \ac{ue}-side \ac{rse} has no access to them.

\subsection{From Traces to Conversations}
Layer-3 messages are sets of \acp{ie}, where each \ac{ie} is a key-value pair whose value may itself be an \ac{ie}. To be transmitted, those sequences are encoded as bit strings according to \acs{asn1} encoding rules. Before an \ac{rse} sees a message, the bit string is decoded and converted into human-readable JSON, which preserves the nested structure of the message while facilitating debugging.

Each captured session (exchange of \ac{rrc} messages between a \ac{ue} and a \ac{gnb}, from the connection request to the release or handover) is then recast as a multi-turn conversation as shown in the example of Figure~\ref{fig:pipeline}. For a \ac{gnb}-side \ac{rse}, the input at a given turn is the history of \ac{rrc} messages exchanged (uplink and downlink) together with the external protocol messages that preceded the decision; the output is the downlink \ac{rrc} message that the \ac{gnbcucp} actually sent. A \ac{ue}-side \ac{rse} is trained symmetrically on the same traces, with the roles of uplink and downlink reversed, but without access to the external protocol messages. This question-answer framing with whole history brings sufficient context to train and use an \ac{rse} via next-token prediction without requiring hand-written state machines.
 
\section{Compact Sequence Engines: Fine-Tuning versus Training from Scratch}
\label{sec:models}
\subsection{Compact Sequence Engines}
Cloud-hosted, extremely large \acp{llm} are a poor fit for AI-RAN. Layer-3 processing time budgets range from \qty{10}{ms} to \qty{2}{s}. For instance, once the \ac{ue} sends an \texttt{RrcSetupRequest}, it waits for an \texttt{RrcSetup} between \qty{100}{ms} and \qty{2}{s} (T300 timer), then the \ac{gnb} expects an \texttt{RrcSetupComplete} within \qty{10}{ms}~\cite[Table 12.1-1]{ETSITS1382026}. Even if the model itself were instantaneous, the network \ac{rtt} between the \ac{gnbcucp} and a distant datacenter would often exceed that budget.

Instead, \acp{rse} could profit from GPUs placed inside the radio access network as envisioned in~\cite{nokiaAIRadioAccess2024} for performing the inference. Nonetheless, those GPUs, when present, will be shared with other AI-for-RAN (from other layers or protocols) and RAN-for-AI (like edge-AI) workloads. Additionally, the constraint is even stricter at the \ac{ue} where mobile terminals have limited compute and a finite battery. 

To keep \acp{rse} small without sacrificing protocol fidelity, we apply three principles to both fine-tuned foundation models and models trained from scratch.

\begin{figure*}
\includegraphics[width=\linewidth]{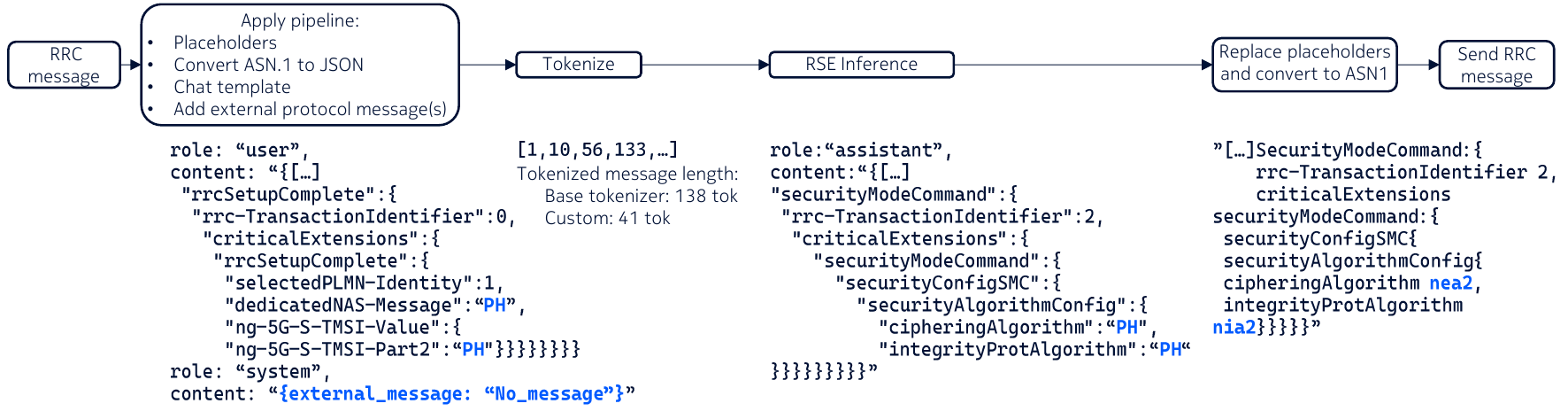}
\caption{Steps performed for generating a response to a received \ac{rrc} message. The message is passed through a pipeline to replace \acp{ie} not needed for \ac{rrc} logic with placeholders, apply a chat template and append external protocol messages (if present). Once the message is generated, the placeholders can be automatically replaced following the network configuration to generate a message that can be encoded and sent.}
\label{fig:pipeline}
\end{figure*}

\paragraph{Custom tokenizers}
Generic tokenizers, like the one used in \cite{liuLLMBasedEmulationRadio2026}, are optimized for natural language but are a poor fit for the vocabulary of \ac{rrc}. Protocol keywords do not occur in ordinary prose, so a standard tokenizer splits each of them into several sub-word fragments, and a message becomes much longer in tokens. Generation time grows with the number of tokens emitted, so a fragmented vocabulary inflates latency. Predicting a keyword fragment by fragment is also harder than predicting it as a whole. Moreover, every fragment is an opportunity to diverge, and therefore a source of hallucination. The vocabulary of \ac{rrc}, however, is fixed by the standard and free of the synonymy of natural language. Recurring keywords can be collected automatically from the traces or the specification and added as atomic tokens, either by extending a pre-trained tokenizer (when fine-tuning) or by defining a dictionary tokenizer (when training from scratch). The effect of this procedure can be seen directly through the example message of Figure~\ref{fig:pipeline} the sequence of tokens to input to the \ac{rse} has a length of 138 tokens with the base tokenizer, whereas using a custom tokenizer augmented with the \ac{rrc} vocabulary reduces the length to 41. Such a reduction is particularly interesting for \texttt{RrcReconfiguration} messages which can easily span thousands of tokens.

\paragraph{Context from external protocol messages}
While some \ac{rrc} messages are responses within \ac{rrc} procedures, many are initiated by radio events, timers, UE behavior, or procedures in other protocols. For example, the \ac{amf} may send an \ac{ngap} \texttt{UeContextReleaseCommand} to the \ac{gnbcucp} requesting release of the \ac{ue}-associated connection. This may cause the \ac{gnb}  to send an \texttt{RrcRelease} to the \ac{ue}. If the release decision originates entirely from \ac{amf}-side state or policy, the preceding \ac{rrc} history may contain little to none information indicating that the release is imminent.


\paragraph{Placeholders}
\label{sec:placeholders}
\Ac{rrc} messages contain critical content that, if incorrect, might strongly deteriorate network \acp{kpi}. As a result, not all \acp{ie} in a message should be produced by the \ac{rse}. We distinguish three kinds of \acp{ie}:
\begin{itemize}
  \item \emph{Opaque containers:} \ac{nas} payloads that the \ac{gnb} merely relays between the \ac{ue} and the core. From the point of view of \ac{rrc} they are arbitrary binary or hexadecimal strings; any attempt to predict their content will fail.
  \item \emph{Exogenous} \acp{ie}: values or structure of \acp{ie} decided outside \ac{rrc} logic, typically provisioned by the operator (cell identifiers, physical-resource configurations, or the ciphering and integrity algorithms advertised in \texttt{SecurityModeCommand}).
  \item \emph{Endogenous} \acp{ie}: values or structure of \acp{ie} produced by the protocol logic itself, such as \texttt{rrc-TransactionIdentifier} in most \ac{rrc} messages or the \texttt{measConfig} structure in \texttt{RrcReconfiguration}.
\end{itemize}
Only the last kind is what the \ac{rse} must learn. Opaque and exogenous values are replaced with placeholders before the \ac{rse} inference is invoked. The placeholders inferred are populated with their corresponding values (provided by the operator's or manufacturer's configuration files). The benefits are several. The \ac{rse} stops wasting capacity and latency on unpredictable or sensitive content and concentrates on the decisions that constitute the protocol: message type, message structure, and protocol-related values. The messages themselves become shorter, which speeds up generation. Most importantly, the boundary of what the \ac{rse} is permitted to decide becomes explicit. This matters for critical \acp{ie}: selecting a ciphering algorithm is an operator decision with security implications, thus the placeholder guarantees that it remains one. While opaque containers are usually clearly labeled in the standard, no such endogenous-exogenous classification exists. As a result, for  Section~\ref{sec:results} we first performed an entropy-based classification, where the low-entropy term are marked as exogenous and high-entropy are marked as endogenous, then the classification was manually reviewed via expert-knowledge assisted by \acp{llm}. 


The full processing pipeline with an example can be seen in Figure~\ref{fig:pipeline}.

\subsection{Fine-Tuning a Foundation Model}
The most direct way to obtain an \ac{rse} is to fine-tune a pre-trained \ac{llm}. Such a model already handles nested, brace-delimited syntax and long-range dependencies, so fine-tuning only has to teach it the protocol logic. Guided by the latency requirement above, we start from the smallest capable candidate rather than the most powerful one: a 0.6-billion-parameter Qwen3 model~\cite{yangQwen3TechnicalReport2025} trained on the conversations presented in Section~\ref{sec:system}.

Considering the highly specific task the \ac{rse} performs, the parameters of the model used for general-purpose language ability are likely largely wasted. That observation raises a natural question: can the model be even smaller if it is designed solely for this task?

\subsection{Training from Scratch}
Fine-tuning leaves little control over the architecture, and therefore over the number of parameters. The vocabulary of \ac{rrc} is comparatively small, and the task is highly structured. Thus, a sequence model trained from scratch can in principle learn only the necessary logic.

The first layer of a sequence model is an embedding that maps each token to a vector. Each token in the vocabulary therefore induces a dense connection into the embedding dimension, and that width propagates to subsequent layers. A small vocabulary and a small embedding dimension thus help maintain a small parameter count. Because the \ac{rrc} vocabulary is defined by the standard and does not contain synonyms, a word-based tokenizer yields both a compact vocabulary and an embedding that can remain small.

A natural choice for the remaining layers would be one or more decoder-only transformer blocks, as in most \acp{llm}. Transformers, however, are a costly match for this workload. Self-attention scales quadratically with sequence length, and a single \ac{rrc} procedure may require a long history of exchanged messages, including contextual information from other control-plane protocols. Some messages, such as \texttt{RrcReconfiguration}, can themselves span thousands of tokens, even with a word-based tokenizer.

Mamba~\cite{guMambaLinearTimeSequence2024a} offers an alternative rooted in selective state-space models. Instead of materializing pairwise token interactions through self-attention, Mamba maintains a compact recurrent state that is updated selectively as a function of the input. The resulting time and memory complexity is linear in the sequence length. A hardware-aware implementation further places intermediate variables in \acl{hbm} or \acl{sram} as appropriate. At inference time, previously computed states can be cached analogously to key-value caching in transformers, so auto-regressive generation need not recompute the full context at every step.

\section{Results}
\label{sec:results}
\subsection{Dataset and evaluation method}
All \acp{rse} presented here are trained on \ac{gnbcucp} traces from over-the-air prototype test networks and structured as multi-turn conversations, as described in Section~\ref{sec:system}. The corpus comprises 5047 independent \ac{gnb}-side sessions, each comprising the \ac{rrc} exchange history together with the contextual external protocol messages (\ac{ngap}, \ac{f1ap}, \ac{e1ap}, \ac{xnap}) that precede a decision. Reciprocally, the corpus contains 4009 \ac{ue}-side sessions (sessions comprising only one \ac{ue}-initiated message and its associated \ac{gnb} response being discarded).

Sessions are randomly split into training, validation, and test sets in a 60/20/20\,\% ratio. In the test set, each multi-turn conversation is transformed into multiple single-turn conversations where the input to the model consists of the whole history up to the message to be generated, with the target message used as ground truth; this isolates per-message accuracy and prevents an early generation error from contaminating later predictions. This makes a total of 14744 messages for the \ac{gnb} and 9775 for the \ac{ue}. The sampling temperature of the \acp{rse} is set to 0. 


\subsection{Metrics}
We evaluate the different \acp{rse} along three axes: generation correctness, syntactic validity, and inference latency.

For correctness, we primarily report \ac{em}, defined as the fraction of generated messages that are exactly identical to the reference. \ac{em} is a demanding metric, but it is well suited to protocol emulation as the impact of a single wrong \ac{ie} is hard to reliably characterize without running a live environment. A single error may have no impact at all or violate the whole protocol logic. We therefore prefer \ac{em} over metrics borrowed from open-ended text generation. For instance, \acs{bleu}~\cite{papineniBleuMethodAutomatic2002a}, which scores $n$-gram overlap, can reach values near maximum even when the message structure is broken or mandatory \acp{ie} are absent, because shared keywords still contribute to the score. Additionally, cosine similarity, as used in the \ac{rse} of~\cite{liuLLMBasedEmulationRadio2026}, is also too permissive: two different message types that share vocabulary (like \texttt{RrcSetup} and \texttt{RrcReconfiguration}) will appear close in the embedding space but will lead to different actions at the receiver.

Syntactic validity is defined as the fraction of outputs that produce a valid \texttt{.json} structure. A high syntactic validity means that generated messages are well-formed and can be parsed; it does not imply that the content matches the reference, which is the role of \ac{em}. 

Latency is reported as the median \ac{ttlt} over the test set, inference being carried out sequentially (batch size of 1).

\subsection{Fine-tuning a compact transformer}
Table~\ref{tab:finetuning} summarizes the main fine-tuning variants. The context window used is \num{25000} token at input. Off-the-shelf (second row), Qwen3-0.6B is unusable as an \ac{rse}: it reproduces the reference message in \qty{2}{\percent} of the cases and only \qty{37}{\percent} of its outputs are parsable, confirming that generic language competence alone does not capture \ac{rrc}. Fine-tuned on the curated dataset with the full pipeline (last row), the same \qty{0.6}{B}-parameter model reaches an \ac{em} of \num{0.82} with a syntactic validity of \num{0.92} on the \ac{gnb} side, at a median \ac{ttlt} of \qty{610}{ms}. This is more than three times faster than the Llama-3.2 1B baseline of~\cite{liuLLMBasedEmulationRadio2026} (\qty{2093}{ms}), yet still more than an order of magnitude above the \qty{10}{ms} regime targeted.

The intermediate rows isolate where that accuracy comes from. A model fine-tuned on \ac{rrc} history alone is capped at \ac{em} \num{0.68}: several downlink messages, such as the \texttt{RrcRelease} that follows an \ac{ngap} \texttt{UeContextReleaseCommand}, are triggered exclusively by external events and are simply not predictable using \ac{rrc} context only. Interleaving external protocol messages lifts \ac{em} to \num{0.79}. 

The same recipe transfers to the \ac{ue} side, where the \ac{rse} reaches \ac{em} \num{0.80} with a syntactic validity of \num{0.99} although it has no access to the external protocol messages present in the dataset. A richer dataset with internal \ac{ue} data would likely further increase performance. Generation is also faster: the median \ac{ttlt} drops from \qty{610}{ms} on the \ac{gnb} side to \qty{320}{ms} on the \ac{ue} side, because uplink messages are generally shorter than downlink ones. 

\begin{table}[]
  \caption{Performance of fine-tuned \acp{rse}.}
  \label{tab:finetuning}
  \centering
  \setlength{\tabcolsep}{3pt}
  \begin{tabular}{lcccc}
    \hline
    Configuration & Side & \ac{em} & Syntactic & Med. \ac{ttlt} \\
    & & & validity & (\unit{ms}) \\
    \hline
    Llama-3.2 1B \cite{liuLLMBasedEmulationRadio2026} \tablefootnote{Metrics on an LTE corpus different from our NR test set with different data preparation and prompt method. We take for reference Llama-3.2 1B (fastest) with FP16 (as we do not quantize our models) with "RRC\_constrain" (best similarity) with full fine-tuning (like our work). \ac{em} was not reported.} & \ac{gnb} & -- & \num{0.96} & \num{2093} \\ 
    Base Qwen3-0.6B\tablefootnote{Median TTLT has not been measured for Qwen0.6B other than the full pipeline} & \ac{gnb} & \num{0.02} & \num{0.37} & --\\
    Qwen3-0.6B, \ac{rrc} context only & \ac{gnb} &  \num{0.68} & \num{0.85} & -- \\
    Qwen3-0.6B, \ac{rrc} and ext. context  & \ac{gnb} & \num{0.79} & \num{0.87} & -- \\ 
    Qwen3-0.6B, full pipeline &\ac{gnb} & \num{0.82} & \num{0.92} & \num{610} \\ 

    Qwen3-0.6B, full pipeline & \ac{ue} & \num{0.80} & \num{0.99} & \num{320} \\ 
    \hline
  \end{tabular}
\end{table}

\subsection{A minimal Mamba-based sequence engine}
Because a \qty{0.6}{B}-parameter pre-trained \ac{llm} still devotes most of its capacity to general-language competence that \ac{rrc} never uses, we train a Mamba model~\cite{guMambaLinearTimeSequence2024a} from scratch on the same dataset using a dictionary-based tokenizer. Results are presented in Table~\ref{tab:mamba}. The context window is \num{80000}, $3\times$ larger than Qwen, thanks to the use of state space models instead of the self-attention matrix.

Firstly, an ablation on data preparation confirms that the same principles govern both backbones. Without placeholders or external protocol messages, a \qty{22}{M}-parameter Mamba model reaches only an \ac{em} of \num{0.20} on the \ac{gnb} side; adding placeholders together with non-\ac{rrc} messages raises \ac{em} to \num{0.87}.

Then, studying different Mamba architectures reveals an accuracy-latency trade-off: a \qty{11}{M}-parameter configuration (embedding dimension 512, four Mamba layers) achieves \ac{em} \num{0.90} with an error-free syntactic validity and a median \ac{ttlt} of \qty{115}{ms}. It is worth noting that it exceeds the accuracy of the fine-tuned Qwen, while using \qty{98}{\percent} fewer parameters and being \num{5} times faster. It is also \num{18} times faster than Llama-3.2 1B. The smallest variants ($256\times1$ and $512\times1$) push the \ac{ttlt} below \qty{50}{ms} but suffer a significant drop in \ac{em} (respectively \num{0.69} and \num{0.73}). Increasing the number of Mamba layers above \num{4} does not seem to further improve the \ac{em} but doubles the latency. Finally, the \ac{ue} side also achieves a higher \ac{em} and a \num{4} times faster \ac{ttlt} compared to the fine-tuned Qwen. 

\begin{table}[]
  \caption{Performance of \acp{rse} with a Mamba backbone trained from scratch. "no Ph." denotes that no placeholders are used. Architecture is noted as embedding dimension $\times$ number of Mamba layers.}
  \label{tab:mamba}
  \centering
  \setlength{\tabcolsep}{3pt}
  \begin{tabular}{lcccccc}
    \hline
    Configuration & Arch. & Side & Params & \ac{em} & Syntactic & Med.\ \ac{ttlt} \\
    & & & & & validity & (\unit{ms}) \\
    \hline
    \ac{rrc} only, no Ph. & $512{\times}8$ &\ac{gnb} & 22M & \num{0.20} & $>$\num{0.999} & \num{291} \\
    Full pipeline & $512{\times}8$ &\ac{gnb}& 18M & \num{0.87} & $>$\num{0.999} & \num{265} \\
    Full pipeline & $512{\times}4$ &\ac{gnb}& 11M &\num{0.90} & \num{1} & \num{115} \\
    Full pipeline & $512{\times}1$ &\ac{gnb}& 6M & \num{0.73} & $>$\num{0.999} & \num{48} \\
    Full pipeline & $256{\times}8$ &\ac{gnb}& 6M & \num{0.87} & $>$\num{0.999} & \num{225} \\
    Full pipeline & $256{\times}4$ &\ac{gnb}& 4M & \num{0.79} & \num{1} & \num{97} \\
    Full pipeline & $256{\times}1$ &\ac{gnb}& 3M & \num{0.69} & $>$\num{0.999} & \num{46} \\
    Full pipeline & $512{\times}4$ &\ac{ue}& 7M & \num{0.85} & \num{1} & \num{81} \\
    \hline
  \end{tabular}
\end{table}

\begin{figure}
\centering
\includegraphics[width=\linewidth]{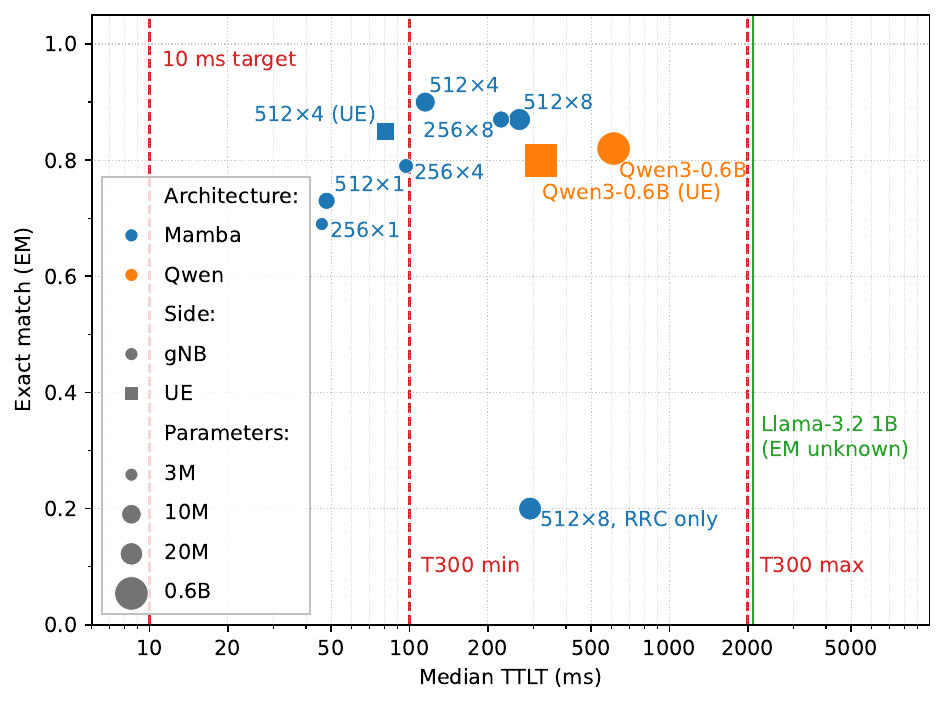}
\caption{Performance evaluation of different \acp{rse}. The solid green vertical line marks the Llama-3.2 1B baseline of~\cite{liuLLMBasedEmulationRadio2026}, whose \ac{em} was not recorded (usage of a less restrictive metric). The dashed line indicates the \qty{10}{ms} Layer-3 latency target and the minimal and maximal values for the T300 as defined in the standard.}
\label{fig:scatter}
\end{figure}

Interestingly, as can be seen in Figure~\ref{fig:scatter}, all \acp{rse} have a median \ac{ttlt} smaller than the maximum T300, suggesting that, with appropriate \ac{ue} configuration, the \ac{gnb}-side \ac{rse} are fast enough for performing the \texttt{RrcSetup} procedure. Additionally, the figure clearly highlights for Mamba-based \ac{rse} the benefit of using custom tokenization, placeholders, and cross-layer context on the \ac{em}. Among backbones, a purpose-built Mamba stack offers a first credible path towards the target of {10}{ms} median response time. 

\section{Conclusion}

Considering the specificities of the \ac{rrc} protocol, pre-trained \acp{llm} are a poor fit to obtain an \ac{rse} capable of running in real time. Fine-tuning improves the performance but only if it is performed jointly with a careful data preparation and the retraining of the tokenizer. Training from scratch a new model is also possible and yields an architecture 55 times smaller achieving higher \ac{em} and faster \ac{ttlt}. Overall, the key take-away of this study is that treating \ac{rrc} as a domain-specific language can yield fast and accurate \acp{rse}. The \qty{10}{ms} target is not met but we can expect optimizations like quantization-aware training to help further reducing the latency without compromising on the accuracy.


Beyond \ac{rrc} itself, \acp{rse} suggest a different way of building the 6G control plane. If Layer~3 behavior can be learned from traces rather than hand-coded from a specification, then the control plane becomes a trainable component that can be updated at the pace of a model release instead of a standardization cycle, and that can easily be specialized to deployments. 

\bibliographystyle{IEEEtran}  
\bibliography{ComsocMagazine}

\end{document}